\documentclass[prb,twocolumn,aps,showpacs,10pt,longbibliography,floatfix]{revtex4-1}

\usepackage[pdftex]{graphicx} 
\usepackage{epstopdf}
\usepackage{verbatim}
\usepackage{color}
\usepackage{subfigure}
\usepackage{amsbsy}
\usepackage{amsmath}
\usepackage{wasysym}

\begin{document}
\title{Magnetic monopole condensation in pyrochlore ice quantum spin liquid: 
application to Pr$_2$Ir$_2$O$_7$ and Yb$_2$Ti$_2$O$_7$}
\author{Gang Chen$^{1,2}$}
\affiliation{${}^1$State Key Laboratory of Surface Physics, 
Center for Field Theory and Particle Physics,
Department of Physics, Fudan University, Shanghai 200433, China}
\affiliation{${}^2$Collaborative Innovation Center of Advanced Microstructures,
Fudan University, Shanghai, 200433, China}

\begin{abstract}
Pyrochlore iridates and pyrochlore ices are two families of materials
where novel quantum phenomena are intertwined with strong spin-orbit coupling,
substantial electron correlation and geometrical frustration.  
Motivated by the puzzling experiments on two pyrochlore systems Pr$_2$Ir$_2$O$_7$ 
and Yb$_2$Ti$_2$O$_7$, we study the proximate Ising orders and the 
quantum phase transition out of quantum spin ice U(1) quantum spin liquid (QSL). 
We apply the electromagnetic duality of the compact quantum electrodynamics 
to analyze the ``magnetic monopoles'' condensation for U(1) QSL. 
The monopole condensation transition represents a unconventional 
quantum criticality with unusual scaling laws.  
It naturally leads to the Ising orders 
that belong to the ``2-in 2-out'' spin ice manifold and generically  
have an enlarged magnetic unit cell. 
We demonstrate that the antiferormagnetic Ising state with the ordering 
wavevector ${\bf Q} = 2\pi(001)$ is proximate to U(1) QSL while the 
ferromagnetic Ising state with ${\bf Q}=(000)$ is not proximate to U(1) QSL. 
This implies that if there exists a direct transition from U(1) QSL 
to the ferromagnetic Ising state, the transition must be strongly first order. 
We apply the theory to Pr$_2$Ir$_2$O$_7$ and Yb$_2$Ti$_2$O$_7$.
\end{abstract}

\date{\today}


\maketitle

\noindent{Pyrochlore} iridates (R$_2$Ir$_2$O$_7$)~\cite{Yanagishima01,Matsuhira07} 
have stimulated a wide interest in recent years, 
and many interesting results, including topological 
Mott insulator~\cite{PesinBalents}, 
quadratic band touching~\cite{PhysRevB.82.085111},
Weyl semimetal~\cite{Wan2011,PhysRevB.85.045124,Gchen2012}, 
non-Fermi liquid~\cite{PhysRevLett.111.206401,PhysRevX.4.041027} 
and so on, have been proposed. 
Among these materials, Pr$_2$Ir$_2$O$_7$ is of particular interest.  
In Pr$_2$Ir$_2$O$_7$, the Ir system remains metallic at low 
temperatures~\cite{Nakatsuji06}. More intriguingly, 
no magnetic order was found except a partial spin freezing 
of the Pr moments due to disorder at very low temperatures 
in the early experiments~\cite{Nakatsuji06,Machida07,Machida09}. 
A recent experiment on different Pr$_2$Ir$_2$O$_7$ samples, 
however, discovered an antiferromagnetic long-range order 
for the Pr moments~\cite{MacLaughlin2015}. 
While most theory works on pyrochlore iridates focused on the Ir
pyrochlores and explored the interplay between the electron correlation 
and the strong spin-orbit coupling of the Ir 5d 
electrons~\cite{PesinBalents,witczak2014correlated,Schaffer2015}, 
very few works considered the influence and the physics of
the local moments from the rare-earth sites that also form 
a pyrochlore lattice~\cite{Gchen2012,PhysRevB.87.125147,Sungbin2013,tian2015field}.   
In this paper, we address the local moment physics in Pr$_2$Ir$_2$O$_7$
and propose that the disordered state of the Pr moments is 
likely to be in the quantum spin ice (QSI) U(1) quantum spin liquid state. 
We explore the proximate Ising order and the confinement transition of QSI and  
argue that Pr$_2$Ir$_2$O$_7$ could be located near 
such a confinement transition.

\begin{figure}[tp]
\centering
\includegraphics[width=7.5cm]{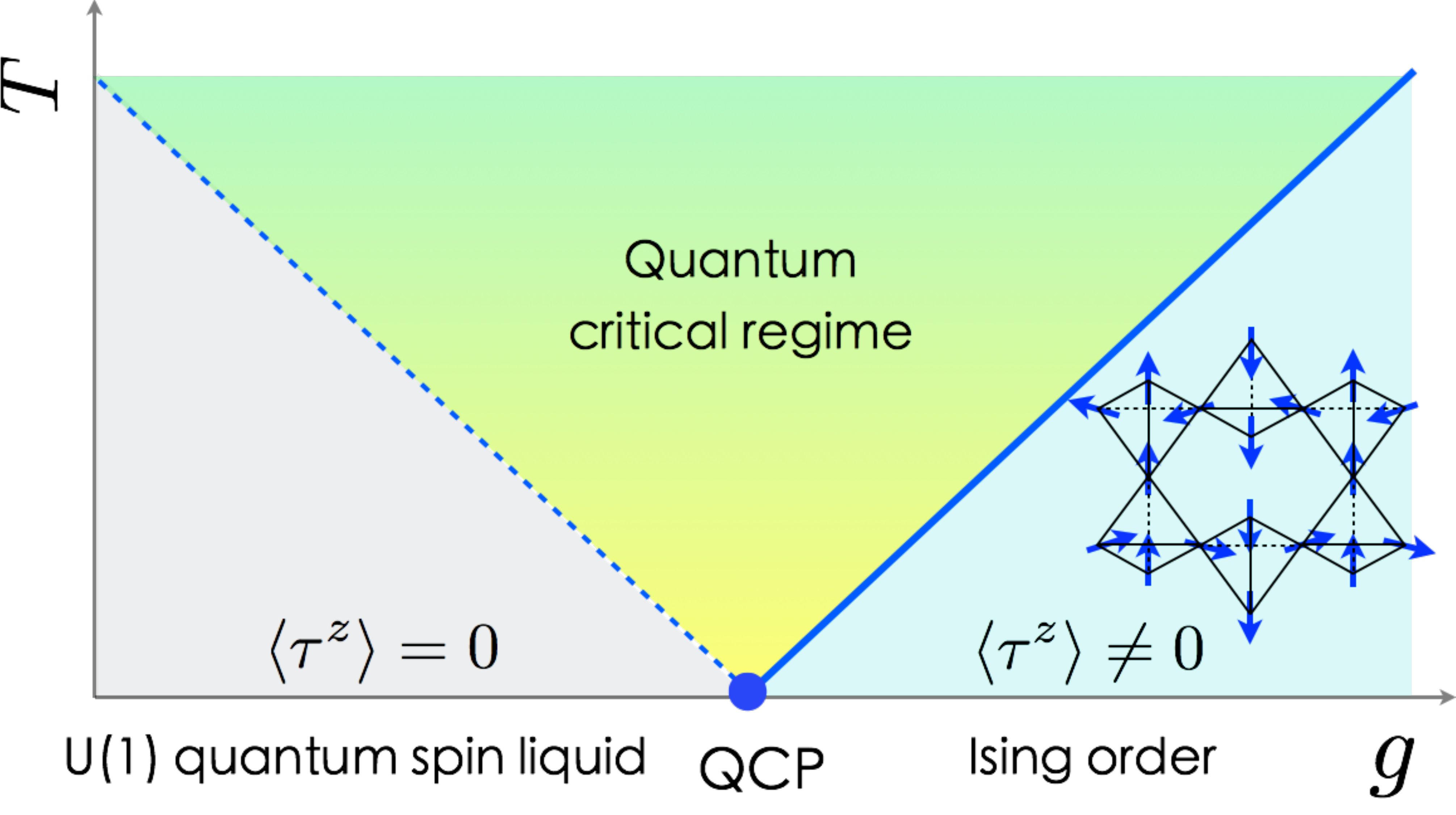}
 \caption{The monopole condensation transition from the QSI U(1) QSL 
 to the proximate antiferromagnetic Ising state. The dashed (solid) line represents 
a thermal crossover (transition). ``$g$'' is a tuning parameter 
that corresponds to the mass of ``magnetic monopole''
(see the discussion in the main text). 
The inset Ising order has an ordering wavevector ${\bf Q} = 2\pi(001)$. 
The Pr moment of Pr$_2$Ir$_2$O$_7$ is likely to be close 
to this quantum critical point (QCP).
}
\label{fig1}
\end{figure}

The QSI U(1) QSL is an exotic quantum phase of matter 
and is described by emergent compact quantum electrodynamics, 
or equivalently, by the compact U(1) lattice gauge theory (LGT) with a
gapless U(1) gauge photon and deconfined spinon 
excitations~\cite{Hermele04,PhysRevLett.98.157204,PhysRevLett.91.167004}. 
Recently several rare-earth pyrochlores with 4f electron local moments 
are proposed as candidates for 
QSI U(1) QSLs~\cite{BalentsSavary,Ross11,Kimura2012,Chang2012,PhysRevLett.108.247210,PhysRevLett.82.1012,PhysRevB.64.224416,PhysRevB.86.075154,Yasui2002,PhysRevLett.109.017201,PhysRevLett.96.177201}. 
In these systems, the predominant antiferromagnetic exchange 
interaction between the Ising components of the local moments 
favors an extensively degenerate 
``2-in 2-out'' spin ice manifold on the pyrochlore 
lattice~\cite{Gingras2001,PhysRevLett.98.157204,Harris1997,Castelnovo12008,RevModPhys.82.53,Gingras2014,BalentsSavary}. 
The transverse spin interaction allows the system to tunnel 
quantum mechanically within the ice manifold, 
giving rise to a U(1) QSL ground state~\cite{PhysRevB.90.214430,RevModPhys.82.53,Gingras2014,Savary12,Sungbin2012,Huang2014}. 
Like Pr$_2$Ir$_2$O$_7$, the experimental results 
on these QSL candidate materials depend sensitively on 
the stoichiometry and the sample preparation~\cite{BalentsSavary}. 
In particular, for the pyrochlore ice system Yb$_2$Ti$_2$O$_7$, 
while some samples remain disordered down to the lowest temperature
and the neutron scattering shows a diffusive scattering~\cite{Ross2009,Ross11}, 
others develop a ferromagnetic order~\cite{Yasui2003,Chang2012,Chang2014,Lhotel2014}. 
This suggests that both the Yb moments in Yb$_2$Ti$_2$O$_7$ and the Pr 
moments in Pr$_2$Ir$_2$O$_7$ could be located near a phase transition 
between a disordered state (that might be a QSI U(1) QSL) 
and the magnetic orders. 

On the theoretical side, the instability of the QSI U(1) QSL 
and the proximate magnetic orders have not been fully explored. 
The early works based on the gauge mean-field approach 
studied the instability by spinon condensation. 
The spinon condensation transition, known as ``Anderson-Higgs transition'', 
generically leads to the transverse spin order 
that is not in the spin ice manifold~\cite{PhysRevB.90.214430}.  
Instead, we here study the proximate Ising spin order and transition 
out of QSI U(1) QSL by condensing the ``magnetic monopoles''
that are topological excitations of the compact U(1) LGT 
for the U(1) QSL~\footnote{The ``magnetic monopole'' used here 
is distinct from the magnetic monopole of Ref.~\onlinecite{Castelnovo12008}. So 
a quotation mark is used. }. The monopole condensation transition 
is the {\sl confinement transition} of the compact U(1) 
LGT~\cite{PhysRevD.19.3682,Savit1980}, 
and the resulting proximate Ising order 
is in the ice manifold and generically breaks the  
translation symmetry. We determine the structure of the 
proximate Ising orders of the QSI U(1) QSL and explain the nature of 
the phase transition from the QSI U(1) QSL to the Ising orders. 

\vspace{0.5cm}

\noindent{\bf Results.}\\
\noindent{\small\bf Compact QED and electromagnetic duality.} 
Even though more complicated realistic Hamiltonians are available 
for effective spin-1/2 moments on the pyrochlore 
lattice~\cite{Savary12,Sungbin2012,Huang2014}, 
it is known that the spin-1/2 XXZ model~\cite{Hermele04},
\begin{equation}
H = \sum_{\langle ij \rangle} \big[  J_{z}^{\phantom\dagger} \tau^z_i \tau^z_j 
{- J_{\perp}^{\phantom\dagger}} (\tau^+_i \tau^-_j + \tau^-_i \tau^+_j) 
 \big],
\end{equation}
in the perturbative regime ($|J_{\perp}|/J_z \ll 1$) 
already captures the {\it universal} properties of the QSI U(1) QSL. 
Here $J_z >0$, $\tau^{\pm}_i \equiv \tau^x_i \pm i \tau^y_i$,
and $\tau^z_i$ is defined along the local $\langle 111 \rangle$ 
direction of each pyrochlore site.  
In the perturbative regime, the third order degenerate perturbation
yields a ring exchange model~\cite{Hermele04},
\begin{eqnarray}
H_{\text{ring}} = -\sum_{\hexagon_p}\frac{K}{2}  
( \tau^+_1 \tau^-_2 \tau^+_3 \tau^-_4 \tau^+_5 \tau^-_6 + h.c.),
\label{ringexchange}
\end{eqnarray} 
where $K= 24 J_{\perp}^3/J_z^2$ and ``1,$\cdots$,6'' 
are 6 sites on the perimeter of the elementary hexagons 
(``$\hexagon_p$'') of the pyrochlore lattice. 

To map the ring exchange model to the compact U(1) LGT, 
one introduces the lattice vector gauge fields as~\cite{Hermele04}
\begin{eqnarray}
E_{{\bf r}{\bf r}'} & \equiv & \tau^z_i+\frac{1}{2}, 
\\
e^{ \pm i A_{{\bf r}{\bf r}'}} & \equiv & \tau^{\pm}_i,
\end{eqnarray} 
where the pyrochlore site $i$ resides on the center 
of the nearest-neighbor diamond link $\langle {\bf r}{\bf r}' \rangle$,
and ${\bf r}$ (${\bf r}'$) is on the I (II) sublattice of the diamond lattice
that is formed by the centers of the tetrahedra. 
Moreover, $E_{{\bf r}{\bf r}'}= -E_{{\bf r}'{\bf r}}, 
A_{{\bf r}{\bf r}'}= -A_{{\bf r}'{\bf r}}$ and 
$[ E_{{\bf r}{\bf r}'}, A_{{\bf r}{\bf r}'}]=i$. 
Here $E_{{\bf r}{\bf r}'}$ ($A_{{\bf r}{\bf r}'}$) 
is integer valued ($2\pi$ periodic). 
With this transformation, $H_{\text{ring}}$ is transformed into  
the compact U(1) LGT on the diamond lattice formed 
by the centers of the tetrahedra,
\begin{eqnarray}
H_{\text{LGT}} = \sum_{\langle {\bf r}{\bf r}' \rangle} 
 \frac{U}{2}(E_{{\bf r}{\bf r}'} - \frac{\epsilon_{\bf r}}{2})^2 
                - \sum_{\hexagon_d}  K\cos (curl\, A), 
\label{eq2}
\end{eqnarray}
where we have added the electric field term with the stiffness $U$, 
$\epsilon_{\bf r}= +1 (-1)$ for ${\bf r} \in$ I (II) sublattice,
and the lattice curl ($curl\, A \equiv \sum_{{\bf r} {\bf r}' 
\in \hexagon_d} A_{{\bf r}{\bf r}'}$) defines the internal 
magnetic field $B$ through the center of the diamond hexagon ($\hexagon_d$). 
In the large $U$ limit, the microscopic $\tau^z = \pm 1/2$ is recovered. 
Although the actual values of $U$ and $K$ in the low energy description of
U(1) QSL are renormalized from the perturbative results, $H_{\text{LGT}}$,
that captures the universal properties of U(1) QSI QSL~\cite{Hermele04},  
is the starting point of our analysis below.

\begin{figure}[tp]
\centering
\includegraphics[width=8.5cm]{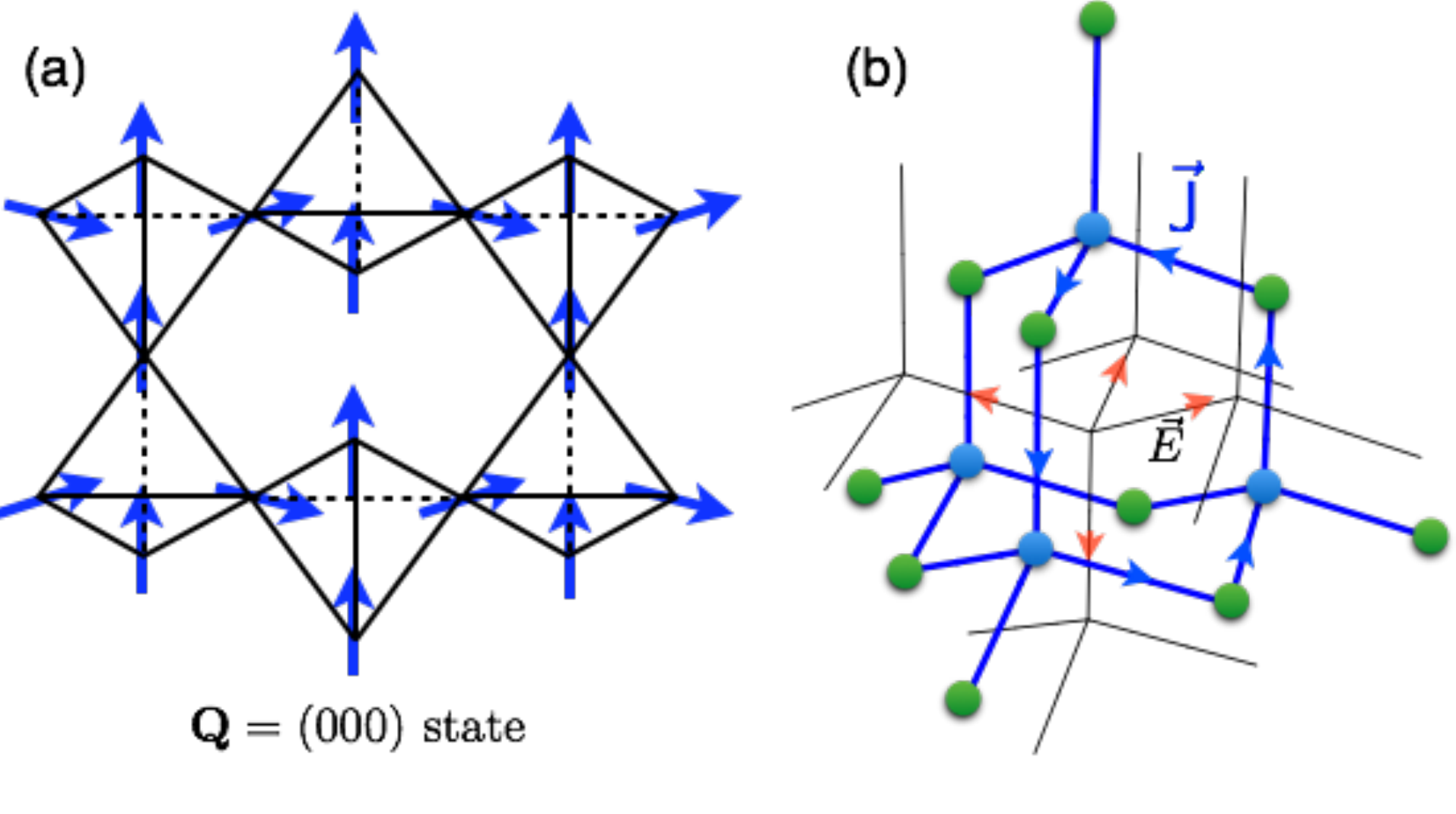}
 \caption{(a) The ${\bf Q} =(000)$ ferromagnetic state. 
 (b) The diamond lattice (in thin black) and the dual diamond lattice (in thick blue). 
 The monopole loop current ($\vec{\mathsf{J}}$) on the hexagon of the dual diamond lattice
  gives rise to the electric field ($\vec{E}$) on the link of the diamond lattice
 via the right hand's rule.  
}
\label{fig2}
\end{figure}

``Magnetic monopoles'' are topological defects of the U(1) gauge field and
carry the magnetic charge. To describe the magnetic transition from U(1) QSL
via the monopole condensation, it is inconvenient to work with 
the field variables in Eq.~\eqref{eq2} because the monopole
variable is not explicit~\cite{Hermele04}. Instead, we apply 
the electromagnetic duality~\cite{PhysRevLett.47.1556,Savit1980,PESKIN1978122,Hermele04,Bergman2006,Motrunich2005,Balents20072635}
to reformulate the compact U(1) LGT Hamiltonian 
and make the monopole explicit. We first introduce an 
integer-valued dual U(1) gauge field 
$a_{\boldsymbol{\mathsf{r}}\boldsymbol{{\mathsf{r}}'}}$
that lives on the link of the dual diamond lattice 
(see Fig.~\ref{fig2}b) such that 
\begin{equation}
curl\, a \equiv 
\sum_{\boldsymbol{\mathsf{r}}\boldsymbol{{\mathsf{r}}'} \in \hexagon^{\ast}_d} 
a_{\boldsymbol{\mathsf{r}}\boldsymbol{{\mathsf{r}}'}}  
\equiv E_{{\bf r}{\bf r}'} - E^0_{{\bf r}{\bf r}'},
\end{equation}
where ``$\hexagon^{\ast}_d$'' refers to the elementary hexagon on the  
dual honeycomb lattice and the electric field vector $E_{{\bf r}{\bf r}'}$
penetrates through the center of ``$\hexagon^{\ast}_d$''. 
Here the serif symbols $\boldsymbol{\mathsf{r}}, \boldsymbol{\mathsf{r}}'$
label the dual diamond lattice sites. 
We have introduced a background electric field distribution 
$E^0_{{\bf r}{\bf r}'}$ that takes care of the background 
charge distribution due to the ``2-in 2-out'' ice rule. 
Each state in the spin ice manifold corresponds to a background 
electric field distribution. For our convenience, 
we choose a simple electric field configuration that corresponds to
a uniform ``2-in 2-out'' spin ice state (see Fig.~\ref{fig2}a) 
with 
\begin{eqnarray}
E^0_{{\bf r},{\bf r}+ \epsilon_{\bf r} {\bf e}_0} 
&=& E^0_{{\bf r},{\bf r}+\epsilon_{\bf r}{\bf e}_1} 
=\epsilon_{\bf r},
\\
E^0_{{\bf r},{\bf r}+ \epsilon_{\bf r}{\bf e}_2} 
&=& E^0_{{\bf r},{\bf r}+ \epsilon_{\bf r}{\bf e}_3} 
= 0,
\end{eqnarray}
where ${\bf e}_{\mu}$ ($\mu = 0,1,2,3$) are the four 
vectors that connect the I sublattice sites 
of the diamond lattice to their nearest neighbors. 
In terms of the dual gauge variables, $H_{\text{LGT}}$ 
is transformed into  
\begin{eqnarray}
H_{\text{dual}} &=& \sum_{\hexagon^{\ast}_d} \frac{U}{2} (curl \, a - \bar{E})^2 
- \sum_{ \langle  {\boldsymbol{ \mathsf{r}}},{\boldsymbol{ \mathsf{r}}}' \rangle } 
 K \cos B_{{\boldsymbol{ \mathsf{r}}}{\boldsymbol{ \mathsf{r}}}'},
\label{dual}
\end{eqnarray}
where we have explicitly replaced $curl\, A$ with the magnetic field vector 
$B_{ {\boldsymbol{ \mathsf{r}}}{\boldsymbol{ \mathsf{r}}}' }$ that lives on the 
link $\langle \boldsymbol{\mathsf{r}}{\boldsymbol{ \mathsf{r}}}'\rangle$ of the dual 
diamond lattice and is conjugate to the dual gauge field $a$ with 
$[ B_{\boldsymbol{\mathsf{r}} \boldsymbol{\mathsf{r}}'}, 
a_{\boldsymbol{\mathsf{r}} \boldsymbol{\mathsf{r}}'}] = i.$
In Eq.~\eqref{dual}, we have introduced the electric field $\bar{E}$
that combines both the background electric field distribution $E^0$
and the offset in Eq.~\eqref{eq2} with 
\begin{eqnarray}
\bar{E}_{{\bf r},{\bf r}+\epsilon_{\bf r}{\bf e}_{\mu}}
= {E}_{{\bf r},{\bf r}+\epsilon_{\bf r}{\bf e}_{\mu}}^0 -  \frac{\epsilon_{\bf r}}{2} .
\end{eqnarray}

Since the dual gauge field $a$ is integer valued, 
the dual Hamiltonian $H_{\text{dual}}$ is difficult to work with. 
Moreover, the ``magnetic monopole'' is implicit in the dual 
gauge field configuration. To make the monopole explicit, 
we follow the standard procedure~\cite{Hermele04}, 
first relax the integer valued constraint of the dual gauge field 
by introducing $\cos 2\pi a$ and then insert the monopole operators. 
The resulting dual theory is described by the magnetic monopoles 
minimally coupled with the dual U(1) gauge field on the dual diamond lattice, 
\begin{eqnarray}
H_{\text{dual}} &=& \sum_{\hexagon^{\ast}_d} \frac{U}{2} (curl \, a - \bar{E})^2 
-\sum_{{\boldsymbol{ \mathsf{r}}},{\boldsymbol{ \mathsf{r}}}'} 
K \cos B_{{\boldsymbol{ \mathsf{r}}}{\boldsymbol{ \mathsf{r}}}'}
\nonumber \\
&& -\sum_{\langle {\boldsymbol{ \mathsf{r}}},{\boldsymbol{ \mathsf{r}}}' \rangle} 
 t \cos (\theta_{\boldsymbol{\mathsf{r}}} - \theta_{\boldsymbol{\mathsf{r}}'} 
 + 2\pi {a_{{\boldsymbol{\mathsf{r}}}{\boldsymbol{\mathsf{r}}}'}}),
\label{dualH}
\end{eqnarray}
where $e^{-i\theta_{\boldsymbol{\mathsf{r}}}}$ 
($e^{i\theta_{\boldsymbol{\mathsf{r}}}}$) 
creates (annihilates) the ``magnetic monopole'' 
at the dual lattice site $\boldsymbol{\mathsf{r}}$.

\vspace{0.5cm}

\begin{figure}[tp]
\centering
\includegraphics[width=4.8cm]{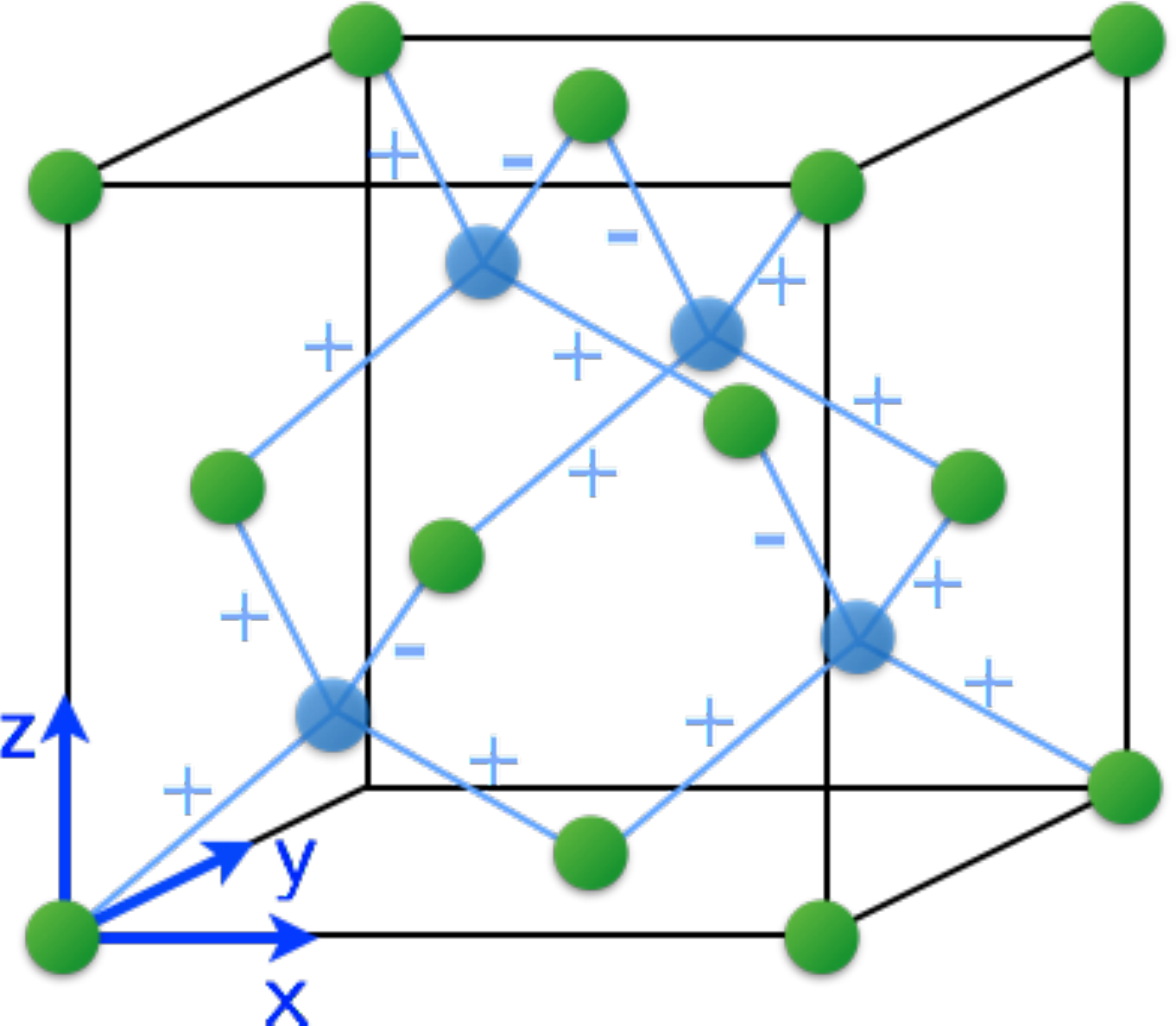}
 \caption{The dual diamond lattice and the assignment of the 
 gauge potential 
 $ e^{- i 2\pi \bar{a}_{{\boldsymbol{ \mathsf{r}}}{\boldsymbol{ \mathsf{r}}}'}} $ 
 on the nearest neighbor links.}
\label{fig3}
\end{figure}

\noindent{\small\bf Monopole condensation and proximate Ising order.}
In the dual gauge Hamiltonian of Eq.~\eqref{dualH}, 
as the monopole hopping increases, the monopole gap decreases. 
When the monopole gap is closed, the monopole is condensed.  
In the confinement phase, the $E$ field develops a static distribution, 
the $B$ field (the $a$ field) is strongly (weakly) fluctuating. 
Therefore, it is legitimate to first ignore the $a$ field fluctuation,
then study the monopole band structure,
and condense the monopoles at the minimum of the monopole band
for the confinement phase~\cite{Motrunich2005,Bergman2006}. 
In such a dual gauge mean-field-like treatment,
the ``$U$'' term in the Hamiltonian enforces $curl \, \bar{a} = \bar{E}$,
which is solved to fix the gauge for the dual gauge field. Here 
we set the dual gauge field to its static component $\bar{a}$. 
The electric field distribution $\bar{E}$ 
turns into the dual gauge flux experienced 
by the ``magnetic monopoles'' in the dual formulation. 
As $\bar{E}$ takes $\pm {\epsilon_{\bf r}}/{2}$, 
it leads to $\pi$ flux of the dual gauge field 
through each elementary hexagon on the dual diamond lattice. 
As it is shown in Fig.~\ref{fig3}, we fix the gauge by setting 
$\bar{a}_{\boldsymbol{\mathsf{r}},\boldsymbol{\mathsf{r}}
+ \boldsymbol{\mathsf e}_{\mathsf{\mu}}} = 
\xi_{\mathsf{\mu}} (\boldsymbol{\mathsf{q}}\cdot 
\boldsymbol{\mathsf{r}})$, 
where $\boldsymbol{\mathsf{r}} \in $ I sublattice of the dual diamond lattice, 
$\boldsymbol{\mathsf e}_{\mathsf{\mu}}$ ($\mathsf{\mu}=\mathsf{0,1,2,3}$) refer to the 
four nearest-neighbor vectors of the dual diamond lattice,
$(\xi_{\mathsf{0}},\xi_{\mathsf{1}},\xi_{\mathsf{2}},\xi_{\mathsf{3}}) = (0110)$ 
and $\boldsymbol{\mathsf{q}}=2\pi(100)$.

In the presence of the background flux, the monopole nearest-neighbor 
hopping model on the dual diamond lattice is given by
\begin{equation}
H_m = - \sum_{\langle {\boldsymbol{ \mathsf{r} }}, {\boldsymbol{ \mathsf{r}} }'\rangle} 
  t \, 
 e^{- i 2\pi \bar{a}_{{\boldsymbol{ \mathsf{r}}}{\boldsymbol{ \mathsf{r}}}'}^{\phantom\dagger}} 
 \Phi^{\dagger}_{\boldsymbol{ \mathsf{r}}}  
 \Phi^{\phantom\dagger}_{{\boldsymbol{ \mathsf{r}}}'} ,
\end{equation}
where we have introduced $\Phi_{\boldsymbol{ \mathsf{r}}}
\equiv e^{i \theta_{\boldsymbol{ \mathsf{r}}}}$ 
(with $|\Phi_{\boldsymbol{ \mathsf{r}}}| \equiv 1 $).
The dispersion of the lowest monopole band is given by
\begin{equation}
\Omega_{ \boldsymbol{\mathsf{k}} } = - |t| [
4 + 2 ( 3 + c_{\mathsf{x}}  c_{\mathsf{y}} 
- c_{\mathsf{x}} c_{\mathsf{z}} 
+ c_{\mathsf{y}} c_{\mathsf{z}} 
)^{1/2}]^{1/2},
\end{equation}
where $c_{\mathsf{\mu}} = \cos {\mathsf{k}_{\mu}}$ ($\mu = \mathsf{x,y,z}$). 
The degenerate minima of the lowest band form several lines 
of momentum points in the Brioullin zone. 
One such degenerate line is along the $[001]$ direction of the Brioullin zone
and the minimum energy is $-2\sqrt{2} |t| $. 
Other degenerate lines are readily obtained by the symmetry operations. 
The line degeneracy of the band minima is a consequence 
of the background flux that frustrates the monopole hopping. 
These continuous degeneracies are accidentical and 
are not protected by symmetry. It is expected that 
the further neighbor monopole hopping 
or monopole interactions should lift these degeneracies.

Because of the background flux, the lattice symmetry in $H_m$
is realized projectively, known as projective symmetry group (PSG)~\cite{WenPSG}. 
We use PSG to generate the further neighbor monopole hoppings~\cite{supple}, 
but do not find obvious degeneracy breaking. Instead, 
the line degeneracy immediately gets lifted if we impose the 
unimodular constraint of the monopole field 
($|\Phi_{\boldsymbol{\mathsf{r}}}| = 1$). 
This unimodular constraint, that originates from the repulsive 
interaction between monopoles,
suppresses the magnitude fluctuation of the monopole fields. 
For the degenerate minima along the [001] direction, 
the unimodular requirement selects the monopole configurations at
two equivalent momenta 
\begin{equation}
\boldsymbol{\mathsf{k}}_1 = (0,0,\pi),
\quad
\boldsymbol{\mathsf{k}}_2 = (0,0,-\pi),
\end{equation}
and the corresponding monopole configurations are 
\begin{eqnarray}
&&\left\{
\begin{array}{lll}
\boldsymbol{\mathsf{r}} & \in \text{I},& \varphi_1 (\boldsymbol{\mathsf{r}}) = 
(\frac{1 + i}{2}  + \frac{1 - i}{2} e^{i 2 \pi \mathsf{x}} ) e^{i \pi \mathsf{z}} , 
\\
\boldsymbol{\mathsf{r}} & \in \text{II},& \varphi_1 (\boldsymbol{\mathsf{r}}) 
= e^{i \pi \mathsf{z}},
\end{array}
\right. 
\\
&&\left\{
\begin{array}{lll}
\boldsymbol{\mathsf{r}} & \in \text{I}, & \varphi_2 (\boldsymbol{\mathsf{r}}) = 
(\frac{i+1}{2}  + \frac{i-1}{2} e^{i 2 \pi \mathsf{x}} ) e^{-i \pi \mathsf{z}} , 
\\
\boldsymbol{\mathsf{r}} & \in \text{II}, & \varphi_2 (\boldsymbol{\mathsf{r}}) = 
i e^{-i \pi \mathsf{z}},
\end{array}
\right. 
\end{eqnarray}
where $\varphi_a$ refers to the monopole configuration at the momentum 
$\boldsymbol{\mathsf k}_a$. From the above results, we 
use the PSG transformations and generate in total twelve 
symmetry equivalent solutions.

After the unimodular constraint is enforced, the monopoles are condensed 
at only one of the equivalent solutions,  
the spinons are confined and the system develops an Ising order. 
Although the Ising order is induced by the monopole condensation, 
as monopoles are emergent particles 
and are not gauge invariant, the physical property of 
the monopole condensate is encoded in the gauge invariant monopole bilinears.  
Again, symmetry is a powerful tool to establish the relation between
the spin density $\tau^z$ and the monopole bilinears. 
The candidate monopole bilinears are the monopole density 
and the monopole current. Although the monopole density ($\Phi^{\dagger}\Phi$) 
transforms in the same way as the spin density ($\tau^z$) 
under the space group symmetry, they behave oppositely 
under the time reversal. 

As for the monopole current, from the Maxwell's 
equations, the loop integral of monopole current is the electric flux 
through the plaquette enclosed by the loop 
(see Fig.~\ref{fig2}b)~\cite{Bergman2006,Motrunich2005}. We have 
\begin{eqnarray}
\tau^z_i &\sim & E_{{\bf r}{\bf r}'} \sim 
\sum_{\boldsymbol{\mathsf{r}}\boldsymbol{\mathsf{r}}' 
\in \hexagon^{\ast}_d} 
\mathsf{J}_{\boldsymbol{\mathsf{r}}\boldsymbol{\mathsf{r}}'},
\label{mag}
\end{eqnarray}
where the pyrochlore site $i$ is the center of the 
elementary honeycomb $\hexagon^{\ast}_d$ on the dual diamond lattice, and
$\mathsf{J}_{\boldsymbol{\mathsf{r}}\boldsymbol{\mathsf{r}}'} \equiv
i ( \langle \Phi^{\dagger}_{\boldsymbol{\mathsf{r}}} \rangle 
\langle \Phi_{\boldsymbol{\mathsf{r}}'} \rangle
e^{- i \bar{a}_{\boldsymbol{\mathsf{r}}\boldsymbol{\mathsf{r}}'  }}
- h.c. )$ defines the monopole current.  
Here $\langle \Phi_{\boldsymbol{\mathsf{r}}}\rangle$ is the 
expectation value of the monopole field that 
is taken with respect to one of the equivalent solutions. 
In the inset of Fig.~\ref{fig1}, 
we depict the spin density distribution of the 
monopole condensate at $\boldsymbol{\mathsf{k}}_1$. 
The resulting Ising order in the confinement phase 
is an antiferromagnetic state with an ordering 
wavevector ${\bf Q} = 2\pi(001)$, and the four spins 
on each tetrahedron obey the ``2-in 2-out'' ice rule. 
This Ising state breaks the translation symmetry by 
doubling the crystal unit cell. 

The translation symmetry breaking of the proximate magnetic state
is a generic phenomenon. The background 
gauge flux, due to the ``2-in 2-out'' rule, 
shifts the minimum of the monopole band to finite momenta. 
Once the monopole is condensed at the finite momentum, 
the resulting proximate Ising order necessarily breaks the translation symmetry. 
If, however, the ferromagnetic Ising order with ${\bf Q}=(000)$ in Fig.~\ref{fig2}a, 
preserves the translation symmetry and borders with the QSI U(1) QSL, 
the transition beween this ferromagnetic Ising
order and U(1) QSL must be strongly first order. 
In the Method, we write down simple models that do not have a
sign problem for quantum Monte Carlo simulation.
The models can realize both the ferromagnetic and antiferromagnetic Ising orders
and allow the careful numerical study of the phase transitons out
of the QSI U(1) QSL.

\vspace{0.5cm}

\begin{figure}[tp]
\centering
\includegraphics[width=5cm]{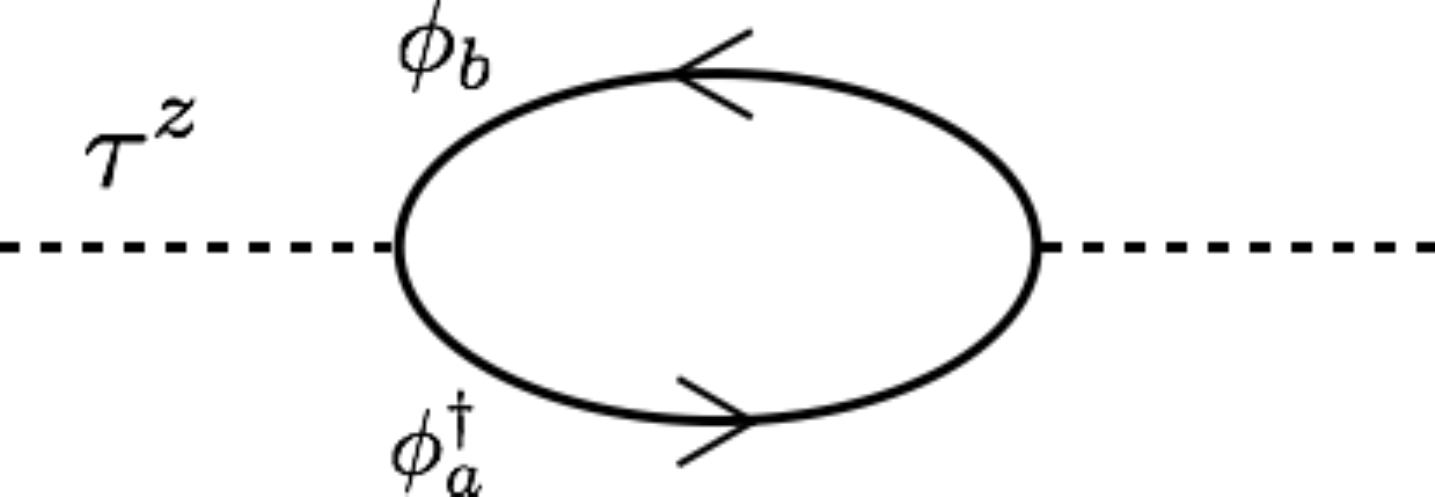}
 \caption{The bubble diagram of the ``magnetic monopole''.}
\label{fig4}
\end{figure}

\noindent{\small\bf Critical theory of monopole condensation.} The monopole 
interaction in the confinement phase selects twelve equivalent monopole 
condensates that correspond to twelve symmetry equivalent Ising orders. 
In the vicinity of the monopole condensation transition, 
the monopole condensate and the gauge fields fluctuate strongly. 
We thereby carry out a Landau-Ginzburg-Wilson expansion 
of the action in terms of the monopole condensate and gauge field 
in the vincinity of the phase transition. 
We introduce the slowly-varying monopole fields $\phi_a$ via the expansion
\begin{eqnarray}
\Phi_{\boldsymbol{\mathsf{r}}} = \sum_{a=1}^{12} 
\varphi_a(\boldsymbol{\mathsf{r}}) \, \phi_a,
\end{eqnarray}
where $\varphi_a (\boldsymbol{\mathsf{r}})$ ($a=1,\cdots,12$) 
are the twelve discrete monopole modes that span the 
ground state manifold of the monopole condensate.
With the monopole PSG, we generate the symmetry allowed effective 
action for the monopole condensation transition, 
\begin{eqnarray}
L &=& \sum_a \big[|(\partial_{\mu} - i \tilde{a}_{\mu} ) \phi_a|^2
+ m^2 |\phi_a|^2 \big]+ \frac{{F_{\mu\nu}}^2}{2} 
\nonumber 
\\
&& + u_0 (\sum_a |\phi_a|^2)^2 + u_1 \sum_{a\neq b} |\phi_a|^2 |\phi_b|^2 + \cdots,
\label{action}
\end{eqnarray}
where we have restored the gauge field fluctuation by coupling the $\phi_a$ fields
to the fluctuating part of the dual U(1) gauge field $\tilde{a}_{\mu}$, 
$\frac{1}{2} {{F_{\mu\nu}}^2}$ is the Maxwell term with 
$F_{\mu\nu} \equiv \partial_{\mu} \tilde{a}_{\nu} -\partial_{\nu} \tilde{a}_{\mu}$, 
 ``$\cdots$'' contains further anisotropic terms that are marginal 
 for the critical properties, $m$ is the mass
of the monopole and is set by the band gap of the monopole band structure. 
The effective action in Eq.~\eqref{action} is a standard 
multi-component Ginzburg-Landau theory in 3+1D
that is the upper critical dimension of the theory. 
One expects the phase transition of this theory to be governed 
by a Gaussian fixed point or belong to a weakly first order transition 
driven by fluctuations~\cite{Halperin1974,Bergman2006,Motrunich2005,Balents20072635,PhysRevB.71.144508}. Both possibilities suggest that the mean-field treatment of 
the phase transition should be sufficient for a rather wide 
range of length scales. In a mean-field description, the  
monopole field correlator at the critical point (with the monopole mass $m=0$) is 
\begin{equation}
\langle \phi_a^{\dagger} (\boldsymbol{\mathsf{k}},\mathsf{\omega}) 
\phi_b^{\phantom\dagger} (\boldsymbol{\mathsf{k}},\mathsf{\omega}) \rangle 
\sim \frac{\delta_{ab}}{\boldsymbol{\mathsf{k}}^2 + \mathsf{\omega}^2}.
\end{equation}
According to Eq.~\eqref{mag}, the spin susceptibility at the ordering wavevector 
${\bf Q}$ is simply given by the bubble diagram of monopole fields (see Fig.~\ref{fig3}) 
and is thus logarithmically divergent at low temperatures with 
\begin{equation}
\chi({\bf Q}) \sim \ln  \frac{1}{T} .
\end{equation}
Such a weak divergence is a unique property of the monopole condensation transition
that is a non-Landau-Ginzburg-Wilson transition. 
For a conventional magnetic transition, 
one would instead have a power-law divergence for the corresponding susceptibility. 
Here, the Ising order is a consequence of the monopole condensation. 
The condensed monopole is the primary order, and the induced 
Ising order is secondary and is thus a perfect example of the subsidiary order~\cite{berg2009charge,PhysRevX.4.031017}.

The monopole mass controls the phase transition and is parameterized as the 
parameter $g$ with $g \equiv -m^2$ in Fig.~\ref{fig1}. 
In the QSI U(1) QSL phase, the monopole is massive with $m^2 > 0$. 
The low energy physics is then governed by the Maxwell's field theory 
and the emergent gapless gauge photon.  Due to the gapless photon, 
the heat capacity of the system behaves as $C_v \sim T^3$ at low temperatures.
As the system approaches the transition from the QSL side, 
the monopole mass decreases. The gapless monopole at the 
criticality gives an extra $T^3$ contribution to 
the heat capacity. Therefore, one would observe an enhancement of the 
$T^3$ heat capacity as the system approaches the criticality. 
Moreover, if one raises temperatures in the U(1) QSL side, the generic argument 
suggests that there is no thermal phase transition except a crossover
due to the thermal population of the ``magnetic monopoles''. 
The populated monopoles simply create thermal confinement of the spinons
at finite temperatures. 
This crossover temperature is set by the mass of the monopoles. 

When $m^2 < 0$, the monopole is condensed and the system develops 
Ising orders. Since the system breaks time reversal symmetry 
on the ordered side, we should have a finite temperature phase 
transition above which the time reversal symmetry is restored. 
The ordering temperature is also set by the mass of the monopoles.

\vspace{0.5cm}

\noindent{\bf Discussion.}

\noindent{\bf \small The transition and the Ising order in Pr$_2$Ir$_2$O$_7$.}
In Pr$_2$Ir$_2$O$_7$, the Pr$^{3+}$ ion has a 4f$^2$ electron configuration 
and form a non-Kramers' doublet which is represented by
a pseudospin-1/2 operator $\boldsymbol{\tau}$ with  
$\tau^z$ ($\tau^x$, $\tau^y$) odd (even) under time reversal $\mathcal{T}$,
\begin{eqnarray}
\mathcal{T}:&& \quad \tau^z \rightarrow -\tau^z , \label{eqtz}\\
\mathcal{T}:&& \quad \tau^{x,y} \rightarrow \tau^{x,y} \label{eqtxy} .
\end{eqnarray}
In the disordered state, a metamagnetic transition 
is observed {\sl only} for magnetic fields along the $\langle 111\rangle$ 
 direction. This is a clear evidence that the disordered state of 
the Pr moments is fluctuating within the ice manifold~\cite{Machida09}
and the metamagnetic transition is a transition from the ``2-in 2-out'' ice
manifold to the ``3-in 1-out'' manifold. 
Since the local moments in QSI U(1) QSL are fluctuating quantum mechanically 
within the ice manifold, this metamagnetic transition in Pr$_2$Ir$_2$O$_7$
is consistent with our proposal that the disordered state of the Pr moments is 
a QSI U(1) QSL.  

Given the non-Kramers' nature of the Pr moment and its unique 
time reversal symmetry properties in Eqs.~\eqref{eqtz} and \eqref{eqtxy}, 
the magnetic order of the Pr moment must be the Ising order with 
$\langle \tau^z \rangle \neq 0$.
If a non-Kramers doublet local moment system has a QSI U(1) QSL ground state, 
the magnetic transition from this state {\sl must} be the confinement transition
of the compact U(1) LGT because a nonzero $\tau^z$ corresponds to 
the static electric field distribution. 
Remarkably, the Ising order that is found  
in the ordered Pr$_2$Ir$_2$O$_7$ samples~\cite{MacLaughlin2015} 
has an ordering wavevector ${\bf Q} = 2\pi(001)$, 
and this is precisely the proximate Ising state 
that we predict from the confinemet transition. 
This experimental result further supports 
our proposal that the disordered state of the Pr moments in 
Pr$_2$Ir$_2$O$_7$ is a QSI U(1) QSL.

In different samples, different oxygen and Ir contents shift
the Fermi energy of the Ir conduction electrons and thus modify
the Ruderman-Kittel-Kasuya-Yosida (RKKY) interaction between the Pr local moments~\cite{Gchen2012,Sungbin2013,YDLee2016}. 
This is likely to be the microscopic origin of the sample dependence. 
Usually the presence of the conduction electron Fermi surface 
modifies the critical properties of the local moment transition. 
But Pr$_2$Ir$_2$O$_7$ is very special. Due to the quadratic band touching 
of the Ir electrons~\cite{kondo2015quadratic,PhysRevB.82.085111}, 
the Fermi energy is very close to the band touching energy and
the Fermi momentum $|{\bf k}_{\text F}|$ is much smaller than the 
wavevector ${\bf Q}$ of the magnetic order. As a result,  
the particle-hole excitations of the Ir system actually decouple 
from the spin fluctuations of the Pr local 
moments at low energies~\cite{RevModPhys.79.1015}. Therefore, the critical 
properties of the Pr local moments are not modified by the conduction electrons. 

At this stage, it is not clear how close the existing Pr$_2$Ir$_2$O$_7$ samples 
are near the phase transition, therefore, it would be interesting to vary the 
Ir and/or oxygen contents in a continuous fashion, to 
drive the system between disordered and ordered phases and directly 
probe the phase transition. 
It is very useful to focus on the disordered Pr$_2$Ir$_2$O$_7$ samples 
and carry out the inelastic neutron scattering. 
Due to the unique time reversal symmetry properties 
in Eqs.~\eqref{eqtz} and \eqref{eqtxy}, only the Ising 
component of Pr local moment couples with the neutron spin. 
As the $\tau^z$ is identified as the emergent electric field, 
the inelastic neutron scattering directly probes the gauge phonon 
excitation. Because of the quadratic band touching, 
the inelastic neutron spectral intensity corresponding to             
the particle-hole excitations of the Ir electrons 
concentrate near the $\Gamma$ point at low energies,
and it does not mix with the gauge phonon modes that 
are peaked near the ``pinch point'' 
momenta~\cite{PhysRevB.86.075154,Savary12}.

\vspace{0.3cm}

\noindent{\bf \small The transition and the magnetic order in Yb$_2$Ti$_2$O$_7$.}
The magnetic state in the ordered Yb$_2$Ti$_2$O$_7$ samples 
has a ${\bf Q}=(000)$ ferromagnetic order and 
preserves the translation symmetry~\cite{Yasui2003,Chang2012,Chang2014,Lhotel2014}, 
though many early experiments found a disordered state~\cite{Ross2009,Ross11}.
The thermal transition from the high-temperature paramagnet 
to the ferromagnetic one is strongly first
order~\cite{Chang2012,Chang2014,Lhotel2014}.  
Unlike the Pr$^{3+}$ moment, the Yb$^{3+}$ moment 
is a Kramers' doublet with all pseudospin components 
odd under time reversal, thus a direct coupling between 
$\tau^z$ and $\tau^{x,y}$ is allowed.
The magnetic transition out of the QSI U(1) QSL for 
the Kramers' doublet can be either an Anderson-Higgs' 
transition~\cite{Chang2012,Savary12,SavaryPRB} or a confinement transition.

In the Higgs' transition scenario~\cite{Chang2012,Savary12,SavaryPRB}, 
a predominant transverse component is induced at the first order 
transition~\cite{SavaryPRB}, and a small Ising component is induced 
similtaneously via the coupling between $\tau^z$ and $\tau^{x,y}$. 
In the scenario of a confinement transition, however,
a predominant Ising order is expected, and this seems to be case in 
Yb$_2$Ti$_2$O$_7$~\cite{Chang2012,Chang2014,Lhotel2014}.
Moreover, as we have explained, the ${\bf Q}=(000)$ Ising order
is not proximate to the QSI U(1) QSL, and the direct 
transition between them through monopole condensation 
must be strongly first order. The strongly first order 
thermal transition in the ordered Yb$_2$Ti$_2$O$_7$ 
samples can thus be naturally regarded as a finite temperature
extension of the zero-temperature one. 
To differentiate the Higgs' and confinement scenarios in Yb$_2$Ti$_2$O$_7$,
it might be helpful to numerically study the microscopic model~\cite{Ross11}
by varying the transverse component interaction and the
Ising component interaction separately and probe the nature of
transition out of the QSI U(1) QSL.

\vspace{0.5cm}

\noindent{\bf \small Summary.}
To summarize, we have studied the Ising magnetic orders out of 
the QSI U(1) QSL via the ``magnetic monopole'' condensation. 
We find that such a confinement transition
gives rise to the proximate Ising ordered state that breaks 
the translation symmetry. We propose that the puzzling 
magnetic properties of Pr$_2$Ir$_2$O$_7$ and Yb$_2$Ti$_2$O$_7$
can be understood from the ``magnetic monopole'' condensation.
Beyond these two systems, we have argued that the magnetic 
transition out of the QSI U(1) QSL for a non-Kramers doublet 
local moments must be a confinement transition via monopole condensation. 
Since the Tb$^{3+}$ local moment in Tb$_2$Ti$_2$O$_7$ is a non-Kramers' doublet,
it is likely that the sample dependent magnetic order in 
Tb$_2$Ti$_2$O$_7$~\cite{PhysRevB.90.014429}
can be understand as the monopole condensation.

\vspace{0.5cm}
\noindent{\bf Method.} \\
\noindent{\bf\small Pyrochlore and dual diamond lattices.}
Pyrochlore lattice is a corner-shared tetrahedral structure in three dimensions.
The centers of the tetrahedra in the pyrochlore lattice form a diamond lattice.   
The dual lattice of the diamond lattice is also a diamond lattice. For the 
dual diamond lattice, we choose the sites
\begin{eqnarray}
\boldsymbol{\mathsf{d}}_{\mathsf 1} & = & (0,0,0) , \\
\boldsymbol{\mathsf{d}}_{\mathsf 2} & = & \frac{1}{4} (1,1,1),
\end{eqnarray}
to be the reference points of the I and II sublattices, respectively.  
The three lattice vectors of the underlying Bravais lattice are 
\begin{eqnarray}
\boldsymbol{\mathsf{a}}_{\mathsf 1} & = & \frac{1}{2}(0,1,1),
 \\
\boldsymbol{\mathsf{a}}_{\mathsf 2} & = & \frac{1}{2}(1,0,1),
 \\
\boldsymbol{\mathsf{a}}_{\mathsf 3} & = & \frac{1}{2}(1,1,0),
\end{eqnarray}
where we have set the lattice constant to unity. 
 
Each site of the dual diamond lattice is connected by four nearest neighbors.
The four vectors $\boldsymbol{\mathsf{e}}_{\mu}$ that connect the neighboring
sites are given as
\begin{eqnarray}
\boldsymbol{\mathsf{e}}_{\mathsf 0} & = & \frac{1}{4} (1,1,1),   \\
\boldsymbol{\mathsf{e}}_{\mathsf 1} & = & \frac{1}{4} (1,-1,-1), \\ 
\boldsymbol{\mathsf{e}}_{\mathsf 2} & = & \frac{1}{4} (-1,1,-1), \\
\boldsymbol{\mathsf{e}}_{\mathsf 3} & = & \frac{1}{4} (-1,-1,1). 
\end{eqnarray}

\vspace{0.5cm}
\noindent{\bf\small Projective symmetry group.}
Both the pyrochlore lattice and the dual diamond lattice 
share the same space group symmetry Fd$\bar{3}$m. 
The Fd$\bar{3}$m space group involves three lattice translations,
\begin{eqnarray}
\mathsf{T}_{i} : \boldsymbol{\mathsf{r}} \rightarrow 
\boldsymbol{\mathsf{r}} + \boldsymbol{\mathsf{a}}_{i},
\end{eqnarray}
a three-fold rotation,
\begin{eqnarray}
\mathsf{C}_3 : (\mathsf{x},\mathsf{y},\mathsf{z}) \rightarrow 
(\mathsf{z},\mathsf{x},\mathsf{y}),
\end{eqnarray}
a two-fold rotation,
\begin{eqnarray}
\mathsf{C}_2 : (\mathsf{x},\mathsf{y},\mathsf{z}) 
\rightarrow (-\mathsf{x},-\mathsf{y},\mathsf{z}), 
\end{eqnarray}
a mirror reflection, 
\begin{eqnarray}
\mathsf{R}: (\mathsf{x},\mathsf{y},\mathsf{z}) 
\rightarrow (\mathsf{y},\mathsf{x},\mathsf{z}), 
\end{eqnarray}
and an inversion,
\begin{eqnarray}
\mathsf{I}: (\mathsf{x},\mathsf{y},\mathsf{z}) 
\rightarrow (\frac{1}{4}-\mathsf{x},
\frac{1}{4}-\mathsf{y},\frac{1}{4}-\mathsf{z}). 
\end{eqnarray}

The physical spin is defined on the pyrochlore lattice site, 
while the ``magnetic monopoles'' are defined on the dual diamond 
lattice sites. Due to the background gauge flux, 
the space group symmetry is realized projectively in the 
monopole hopping Hamiltonian $H_m$. For each symmetry 
operation, we need to supplement with a U(1) gauge transformation.
Under the symmetry operation $\hat{\mathsf O}$, 
the monopole is transformed as
\begin{equation}
\hat{\mathsf O}: \Phi_{\boldsymbol{ \mathsf{r}}}  \rightarrow  
e^{-i \mathsf{\Theta_{{O}}}(\boldsymbol{ \mathsf{r}})}
\Phi_{\boldsymbol{ \mathsf{r'}}},
\end{equation}
where $\boldsymbol{\mathsf{r'}} = { \mathsf O} ( \boldsymbol{\mathsf{r}} )$
and $e^{-i \mathsf{\Theta_{{\mathsf O}}}(\boldsymbol{ \mathsf{r}})}$ 
is the associated U(1) gauge transformation. 
We have used $\hat{\mathsf O}$ to label the generator 
of the projective symmetry group. 

For our convenience, we introduce the unit cell index $\boldsymbol{\mathsf{n}}$ to 
label the monopole position and define 
\begin{eqnarray}
\eta_{\mathsf 1} (\boldsymbol{\mathsf{n}})  =  \Phi_{\boldsymbol{ \mathsf{r}}} ,
\quad
\eta_{\mathsf 2} (\boldsymbol{\mathsf{n}})  =   \Phi_{\boldsymbol{ \mathsf{r}} +
\boldsymbol{\mathsf{e}}_{\mathsf 0}} ,
\end{eqnarray}
where $\boldsymbol{ \mathsf{r}} =\sum_{j} \mathsf{n}_j 
\boldsymbol{ \mathsf{a}}_j $, and $\eta_{\mathsf 1} 
(\boldsymbol{\mathsf{n}})$ and $\eta_{\mathsf 2} 
(\boldsymbol{\mathsf{n}})$ are monopole operators on 
the I and II sublattices, respectively. 

Here we list the projective symmetry transformation of 
the monopole operators. Under the three lattice translations, 
the monopole operators are transformed as
\begin{eqnarray}
\hat{\mathsf{T}}_{1}:  &\quad&  \eta_1 ({\mathsf{n_x}},{\mathsf{n_y}},{\mathsf{n_z}}) 
\rightarrow    
e^{- i \Theta_{\mathsf T_1} [\boldsymbol{\mathsf{n}}] }
\eta_1 ({\mathsf{n_x} +1},{\mathsf{n_y}},{\mathsf{n_z}}) ,
\\
\hat{\mathsf{T}}_{1}:  &\quad&  \eta_2 ({\mathsf{n_x}},{\mathsf{n_y}},{\mathsf{n_z}}) 
\rightarrow    
e^{- i \Theta_{\mathsf T_1} [\boldsymbol{\mathsf{n}}] }
\eta_2 ({\mathsf{n_x} +1},{\mathsf{n_y}},{\mathsf{n_z}}) ,\\
\hat{\mathsf{T}}_{2}:  &\quad&  \eta_1 ({\mathsf{n_x}},{\mathsf{n_y}},{\mathsf{n_z}}) 
\rightarrow    
e^{- i \Theta_{\mathsf T_2} [\boldsymbol{\mathsf{n}}] }
\eta_1 ({\mathsf{n_x}},{\mathsf{n_y} +1},{\mathsf{n_z}}) ,
\\
\hat{\mathsf{T}}_{2}:  &\quad&  \eta_2 ({\mathsf{n_x}},{\mathsf{n_y}},{\mathsf{n_z}}) 
\rightarrow    
e^{- i \Theta_{\mathsf T_2} [\boldsymbol{\mathsf{n}}] }
\eta_2 ({\mathsf{n_x} },{\mathsf{n_y} + 1},{\mathsf{n_z}}) ,\\
\hat{\mathsf{T}}_{3}:  &\quad&  \eta_1 ({\mathsf{n_x}},{\mathsf{n_y}},{\mathsf{n_z}}) 
\rightarrow    
e^{- i \Theta_{\mathsf T_3} [\boldsymbol{\mathsf{n}}] }
\eta_1 ({\mathsf{n_x}},{\mathsf{n_y}},{\mathsf{n_z} +1 }) ,
\\
\hat{\mathsf{T}}_{3}:  &\quad&  \eta_2 ({\mathsf{n_x}},{\mathsf{n_y}},{\mathsf{n_z}}) 
\rightarrow    
e^{- i \Theta_{\mathsf T_3} [\boldsymbol{\mathsf{n}}] }
\eta_2 ({\mathsf{n_x} },{\mathsf{n_y}},{\mathsf{n_z}+1}) ,
\end{eqnarray}
where 
\begin{eqnarray}
\Theta_{\mathsf{T}_i} [\boldsymbol{\mathsf{n}}] 
= - ( \boldsymbol{\epsilon} \cdot \boldsymbol{\mathsf{n}}) \, \mathsf{{v}}_i 
\end{eqnarray}
and $\boldsymbol{\epsilon} = (1,1,0), \mathsf{\bf{v}} = \pi (0,1,1)$. 

Under three-fold rotation, we have 
\begin{eqnarray}
\hat{\mathsf C}_3: &\quad &  \eta_1 ({\mathsf{n_x}},{\mathsf{n_y}},{\mathsf{n_z}}) 
\rightarrow    
e^{- i \Theta_{\mathsf C_3} [\boldsymbol{\mathsf{n}}] }
\eta_1 ({\mathsf{n_z}},{\mathsf{n_x}},{\mathsf{n_y}}) ,
\\
\hat{\mathsf C}_3: &\quad &  \eta_2 ({\mathsf{n_x}},{\mathsf{n_y}},{\mathsf{n_z}}) 
\rightarrow    
e^{- i \Theta_{\mathsf C_3} [\boldsymbol{\mathsf{n}}] }
\eta_2 ({\mathsf{n_z}},{\mathsf{n_x}},{\mathsf{n_y}}) ,
\end{eqnarray}
where 
\begin{eqnarray}
\Theta_{\mathsf{C_3} } [\boldsymbol{\mathsf{n}}]  
= \boldsymbol{\mathsf{n}} \cdot \mathcal{B} \cdot \boldsymbol{\mathsf{n}} 
+ \boldsymbol{\mathsf{\delta}} \cdot \boldsymbol{\mathsf{n}} 
\end{eqnarray}
with 
\begin{eqnarray}
\mathcal{B} = \frac{\pi}{2} \left[ 
\begin{array}{lll}
1&0&1\\
0&1&1\\
1&1&0
\end{array}
\right]
\end{eqnarray}
and $\boldsymbol{\mathsf{\delta}} = \pi/2 (1,1,0) $. 

Under two-fold rotation, we have 
\begin{eqnarray}
\hat{\mathsf C}_2: &\,\, &  \eta_1 ({\mathsf{n_x}},{\mathsf{n_y}},{\mathsf{n_z}}) 
\rightarrow    
\eta_1 ({\mathsf{n_y}},{\mathsf{n_x}},{-\mathsf{n_x}-\mathsf{n_y}-\mathsf{n_z}}) ,
\\
\hat{\mathsf C}_2: &\,\, &  \eta_2 ({\mathsf{n_x}},{\mathsf{n_y}},{\mathsf{n_z}}) 
\rightarrow    
\eta_2 ({\mathsf{n_y}},{\mathsf{n_x}},{-1-\mathsf{n_x}-\mathsf{n_y}-\mathsf{n_z}}) ,
\end{eqnarray}
where $\Theta_{\mathsf C_2} [\boldsymbol{\mathsf{n}}] =0 $. 

Under the reflection, we have 
\begin{eqnarray}
\hat{\mathsf R}: &\,\, &  \eta_1 ({\mathsf{n_x}},{\mathsf{n_y}},{\mathsf{n_z}}) 
\rightarrow    e^{- i \Theta_{\mathsf R} [\boldsymbol{\mathsf{n}}] }
\eta_1 ({\mathsf{n_y}},{\mathsf{n_x}},\mathsf{n_z}) ,
\\
\hat{\mathsf R}: &\,\, &  \eta_2 ({\mathsf{n_x}},{\mathsf{n_y}},{\mathsf{n_z}}) 
\rightarrow     e^{- i \Theta_{\mathsf R} [\boldsymbol{\mathsf{n}}] }
\eta_2 ({\mathsf{n_y}},{\mathsf{n_x}},\mathsf{n_z}) ,
\end{eqnarray}
where 
\begin{eqnarray}
\Theta_{\mathsf{R} }  [\boldsymbol{\mathsf{n}}] = 
\boldsymbol{\mathsf{n}} \cdot \mathcal{B}' \cdot \boldsymbol{\mathsf{n}} 
+ \boldsymbol{\mathsf{\delta'}} \cdot \boldsymbol{\mathsf{n}} 
\end{eqnarray}
with 
\begin{eqnarray}
\mathcal{B}' = \frac{\pi}{2} \left[ 
\begin{array}{lll}
1&1&0\\
1&1&0\\
0&0&0
\end{array}
\right]
\end{eqnarray}
and $\boldsymbol{\mathsf{\delta'}} = \pi/2 (1,1,0) $. 

Finally, for the inversion symmetry, we have 
\begin{eqnarray}
\hat{\mathsf I}: &\,\, &  \eta_1 ({\mathsf{n_x}},{\mathsf{n_y}},{\mathsf{n_z}}) 
\rightarrow    e^{- i \Theta_{\mathsf I} [\boldsymbol{\mathsf{n}}] }
\eta_2 ({-\mathsf{n_x}},{-\mathsf{n_y}},-\mathsf{n_z}) ,
\\
\hat{\mathsf I}: &\,\, &  \eta_2 ({\mathsf{n_x}},{\mathsf{n_y}},{\mathsf{n_z}}) 
\rightarrow     e^{- i \Theta_{\mathsf I} [\boldsymbol{\mathsf{n}}] }
\eta_1 ({-\mathsf{n_x}},{-\mathsf{n_y}},-\mathsf{n_z}) ,
\end{eqnarray}
where 
\begin{eqnarray}
\Theta_{\mathsf{I} }  [\boldsymbol{\mathsf{n}}] 
= \boldsymbol{\mathsf{\lambda}} \cdot \boldsymbol{\mathsf{n}}
\end{eqnarray}
and $\boldsymbol{\mathsf{\lambda}} = \pi (0,1,0) $. 

\vspace{0.5cm}
\noindent{\bf\small Further neighbor monopole hoppings.}
The general monopole hopping model should be invariant under the PSG transformation. 
We here give an example for the second neighbor monopole hopping 
to illustrate the procedure to determine the hopping parameters. 
The second neighbor connects the lattice sites within the same sublattice. 
Each site has twelve second-neighbor
sites. For the sites in the I sublattice, we consider the 
monopole hopping Hamiltonian,
\begin{eqnarray}
H_m' &=& \sum_{ \boldsymbol{\mathsf{n}} } 
d_1 [ \boldsymbol{\mathsf{n}}]\, \eta_1^\dagger( {\mathsf{n_x}},{\mathsf{n_y}},{\mathsf{n_z}} )
\eta_1^{}( {\mathsf{n_x}+1},{\mathsf{n_y}},{\mathsf{n_z}} ) \nonumber \\
&+& d_2 [ \boldsymbol{\mathsf{n}}]\, \eta_1^\dagger( {\mathsf{n_x}},{\mathsf{n_y}},{\mathsf{n_z}} )
\eta_1^{}( {\mathsf{n_x}},{\mathsf{n_y}+1},{\mathsf{n_z}} ) \nonumber \\
&+& d_3 [ \boldsymbol{\mathsf{n}}]\, \eta_1^\dagger( {\mathsf{n_x}},{\mathsf{n_y}},{\mathsf{n_z}} )
\eta_1^{}( {\mathsf{n_x}},{\mathsf{n_y}},{\mathsf{n_z}+1} ) \nonumber \\
&+& d_4 [ \boldsymbol{\mathsf{n}}]\, \eta_1^\dagger( {\mathsf{n_x}},{\mathsf{n_y}},{\mathsf{n_z}} )
\eta_1^{}( {\mathsf{n_x}},{\mathsf{n_y}-1},{\mathsf{n_z}+1} ) \nonumber \\
&+& d_5 [ \boldsymbol{\mathsf{n}}]\, \eta_1^\dagger( {\mathsf{n_x}},{\mathsf{n_y}},{\mathsf{n_z}} )
\eta_1^{}( {\mathsf{n_x}-1},{\mathsf{n_y}},{\mathsf{n_z}+1} ) \nonumber \\
&+& d_6 [ \boldsymbol{\mathsf{n}}]\, \eta_1^\dagger( {\mathsf{n_x}},{\mathsf{n_y}},{\mathsf{n_z}} )
\eta_1^{}( {\mathsf{n_x}},{\mathsf{n_y}-1},{\mathsf{n_z}+1} )  \nonumber \\
&+& h.c.,
\end{eqnarray}
where $\{ d_i [ \boldsymbol{\mathsf{n}} ] \}$ are the hopping parameters.  
Applying the $\hat{\mathsf{T}}_1$ translation, we compare the transformed Hamiltonian with the 
original Hamiltonian and obtain 
\begin{equation}
d_i [{\mathsf{n_x}},{\mathsf{n_y}},{\mathsf{n_z}}] = d_i [{\mathsf{n_x} -1},{\mathsf{n_y}},{\mathsf{n_z}}].
\end{equation}
Similarly, for the $\hat{\mathsf{T}}_2$ and $\hat{\mathsf{T}}_3$ translations, we have 
\begin{eqnarray}
d_1 [{\mathsf{n_x}},{\mathsf{n_y}},{\mathsf{n_z}}] &=& - d_1[{\mathsf{n_x}},{\mathsf{n_y}-1},{\mathsf{n_z}}] ,\\
d_2 [{\mathsf{n_x}},{\mathsf{n_y}},{\mathsf{n_z}}] &=& - d_2[{\mathsf{n_x}},{\mathsf{n_y}-1},{\mathsf{n_z}}] ,\\
d_3 [{\mathsf{n_x}},{\mathsf{n_y}},{\mathsf{n_z}}] &=& + d_3[{\mathsf{n_x}},{\mathsf{n_y}-1},{\mathsf{n_z}}] ,\\
d_4 [{\mathsf{n_x}},{\mathsf{n_y}},{\mathsf{n_z}}] &=& - d_4[{\mathsf{n_x}},{\mathsf{n_y}-1},{\mathsf{n_z}}] ,\\
d_5 [{\mathsf{n_x}},{\mathsf{n_y}},{\mathsf{n_z}}] &=& - d_5[{\mathsf{n_x}},{\mathsf{n_y}-1},{\mathsf{n_z}}] ,\\
d_6 [{\mathsf{n_x}},{\mathsf{n_y}},{\mathsf{n_z}}] &=& + d_6[{\mathsf{n_x}},{\mathsf{n_y}-1},{\mathsf{n_z}}] ,
\end{eqnarray}
and 
\begin{eqnarray}
d_1 [{\mathsf{n_x}},{\mathsf{n_y}},{\mathsf{n_z}}] &=& - d_1[{\mathsf{n_x}},{\mathsf{n_y}},{\mathsf{n_z}-1}] ,\\
d_2 [{\mathsf{n_x}},{\mathsf{n_y}},{\mathsf{n_z}}] &=& - d_2[{\mathsf{n_x}},{\mathsf{n_y}},{\mathsf{n_z}-1}] ,\\
d_3 [{\mathsf{n_x}},{\mathsf{n_y}},{\mathsf{n_z}}] &=& + d_3[{\mathsf{n_x}},{\mathsf{n_y}},{\mathsf{n_z}-1}] ,\\
d_4 [{\mathsf{n_x}},{\mathsf{n_y}},{\mathsf{n_z}}] &=& - d_4[{\mathsf{n_x}},{\mathsf{n_y}},{\mathsf{n_z}-1}] ,\\
d_5 [{\mathsf{n_x}},{\mathsf{n_y}},{\mathsf{n_z}}] &=& - d_5[{\mathsf{n_x}},{\mathsf{n_y}},{\mathsf{n_z}-1}] ,\\
d_6 [{\mathsf{n_x}},{\mathsf{n_y}},{\mathsf{n_z}}] &=& + d_6[{\mathsf{n_x}},{\mathsf{n_y}},{\mathsf{n_z}-1}] ,
\end{eqnarray}
respectively. Applying the remaining symmeties, we obtain the following 
hopping parameters for the second neighbors,
\begin{eqnarray}
d_1 [{\mathsf{n_x}},{\mathsf{n_y}},{\mathsf{n_z}}] &=&  (-)^{ {\mathsf{n_y}} + {\mathsf{n_z}} } t_2,
\\
d_2 [{\mathsf{n_x}},{\mathsf{n_y}},{\mathsf{n_z}}] &=&  -(-)^{ {\mathsf{n_y}} + {\mathsf{n_z}} } t_2,
\\
d_3 [{\mathsf{n_x}},{\mathsf{n_y}},{\mathsf{n_z}}] &=&   t_2,
\\
d_4 [{\mathsf{n_x}},{\mathsf{n_y}},{\mathsf{n_z}}] &=&  (-)^{ {\mathsf{n_y}} + {\mathsf{n_z}} } t_2,
\\
d_5 [{\mathsf{n_x}},{\mathsf{n_y}},{\mathsf{n_z}}] &=&  (-)^{ {\mathsf{n_y}} + {\mathsf{n_z}} } t_2,
\\
d_6 [{\mathsf{n_x}},{\mathsf{n_y}},{\mathsf{n_z}}] &=&  - t_2.
\end{eqnarray}
With the above procedure, we proceed to generate the further neighbor monopole hoppings
up to the fifth neighbors.

\vspace{0.5cm}
\noindent{\bf\small Monopole condensates.}
We consider the nearest neighbor monopole hopping model. Due to the background flux and the 
gauge choice, the unit cell is fictitiously doubled. In Fig.~\ref{fig3}, we specify the signs of 
the hopping parameters on the dual diamond lattice. The lowest energy spectrum has line degeneracies
in the momentum space. Focusing on the [001] direction in the momentum space, we have the 
following eigenstates for a given ${\mathsf{k_z}}$,
\begin{eqnarray}
\boldsymbol{\mathsf{r}} \in \text{I},\quad  \Phi (\boldsymbol{\mathsf{r}}) & = & \frac{1}{\sqrt{2}}
( e^{i \frac{\mathsf{k_z}}{4}}  + e^{-i \frac{\mathsf{k_z}}{4}} e^{i 2 \pi \mathsf{x}} ) e^{i {\mathsf{k_z}} \mathsf{z}} , 
\\
\boldsymbol{\mathsf{r}}  \in \text{II},\quad \Phi (\boldsymbol{\mathsf{r}}) 
& = & e^{i {\mathsf{k_z}} \mathsf{z}}.
\end{eqnarray}
The monopoles are condensed at these lowest energy momenta. 
To satisfy the unimodular condition for the monpoles, we immediately require 
the monopoles to be condensed at $\mathsf{k_z} = \pm \pi $.

\vspace{0.5cm}
\noindent{\bf\small A sign-problem free model for quantum Monte Carlo simulation.}
Here we propose a simple exchange model that does not have a sign
problem for quantum Monte Carlo (QMC) simulation. This model can realize both 
the ${\bf Q}=2\pi(001)$ order and the ${\bf Q} = (000)$ order. Although both 
Ising orders belong to the spin ice manifold, the former is proximate
to the QSI U(1) QSL via a confinement transition and the latter is not 
(see the main text for the detailed discussion). The model is given as
\begin{eqnarray}
H_1 &=& \sum_{\langle ij \rangle} J_{z}^{\phantom\dagger} \tau^z_i \tau^z_j 
- J_{\perp}^{\phantom\dagger} (\tau^+_i \tau^-_j + h.c.) 
\nonumber \\
&& +  \sum_{\langle \langle\langle ij \rangle\rangle \rangle} 
J_{3z}^{\phantom\dagger} \tau^z_i \tau^z_j ,
\label{suppeq}
\end{eqnarray}
where $J_{3z}$ is the third neighbor Ising exchange. 

We focus our discussion on the case when $J_{\perp} >0$. 
This is precisely the parameter regime where the sign problem 
for QMC is absent.  To be in the spin ice regime, we keep $J_z >0$. 
When $J_{\pm} \ll J_z$ and $J_{3z} \ll J_z$, the ground state is a
QSI U(1) QSL. 
If we fix $J_{\pm}/J_z$ to make the system in the 
QSI U(1) QSL phase, as we gradually increase $|J_{3z}/J_z|$ from 0, 
the system will eventually become ordered. 
Since $J_{3z}$ is the interaction between spins from the same sublattice,
a ferromagnetic $J_{3z}$ would simply favor ${\bf Q}=(000)$, 
even though the four spins on each tetrahedron of the pyrochlore 
lattice obey the ``two-in two-out'' ice rule 
(see Fig.~\ref{fig2}a of the main text).  
Since this ${\bf Q}=(000)$ is not proximate to the U(1) QSL phase,
we expect a {\sl strongly first order} transition as we increase
$|J_{3z}/J_z|$ for a ferromagnetic $J_{3z}$. 

For an antiferromagnetic $J_{3z}$, although the Luttinger-Tisza method 
gives a continuous line degeneracy for the ordering wavevector, the 
Ising constraint immediately select
the collinear order with an ordering wavevector ${\bf Q}=2\pi(001)$. 
As we show in the main text, this Ising order is proximate to 
the U(1) QSL via a monopole condensation transition. 
Therefore, we expect either a continuous transition or 
an extremeley weakly first order transition driven by fluctuations
as we increase $|J_{3z}/J_z|$ for an antiferromagnetic $J_{3z}$. 

In the future, it would be interesting to implement a large scale QMC
simulation of the model in Eq.~\eqref{suppeq} to confirm the 
monopole condensation transition out the QSI U(1) QSL.

Finally, we propose a perturbative version of the model in Eq.~\eqref{suppeq}.
The new model includes the ring exchange on the pyrochlore hexagons and 
the third neighbor Ising exchange and is given as
\begin{eqnarray}
H_2 &=& -\sum_{\hexagon_p}\frac{K}{2}  
( \tau^+_1 \tau^-_2 \tau^+_3 \tau^-_4 \tau^+_5 \tau^-_6 + h.c.)
\nonumber \\
&& +  \sum_{\langle \langle\langle ij \rangle\rangle \rangle} 
J_{3z}^{\phantom\dagger} \tau^z_i \tau^z_j ,
\label{suppeq2}
\end{eqnarray}
and we further restrict the Hilbert space to be the ``2-in 2-out'' ice manifold. 
Therefore, this new Hamiltonian will only act on the states in the ice manifold. 
This perturbative model was already proposed in one perturbative limit of the 
realistic spin model for Yb$_2$Ti$_2$O$_7$ in Ref.~\onlinecite{Savary12}. 
When $|J_{3z}| \ll K $, the ground state of $H_2$ is the QSI U(1) QSL phase.
When $|J_{3z}| \gg K $, the system develops ${\bf Q} = 2\pi(001)$ 
antiferromagnetic order for a positive $J_{3z}$, and 
${\bf Q} = (000)$ ferromagnetic order for a negative $J_{3z}$. Again, 
we expect the transition from the QSI U(1) QSL to the ferromagnetic state
is strongly first order, while the transition to the antiferromagnetic state
is either continuous or extremeley weakly first order.

\bibliography{refs}

\vspace{0.5cm}

\noindent{\bf Acknowledgements.}---I am particularly indebted to  
Leon Balents for the clarification of his early work and the encouragement.
I am especially grateful to M.P.A. Fisher for his emphasis on 
universality in a conversation that inspired me significantly.  
I acknowledge very useful conversation with C. Broholm, J.G. Cheng,
X. Dai, G. Fiete, L.Y. Hung, Y.B. Kim, S.S. Lee, S. Nakatsuji, T. Senthil, F. Wang, 
Z.Y. Weng, F.C. Zhang, and Y. Zhou and an early collaboration 
with M. Hermele. I sincerely thank the hospitality of Oleg Tchnyershov
for inviting me to a pleasant trip at Johns Hopkins University where 
some of the insights were developed. Finally, I thank the hospitality 
of Fuchun Zhang and Yi Zhou for supporting my stay at Zhejiang University 
in April and May 2015 when and where most part of the work was carried out. 
The work is supported by the starting-up fund of Fudan University 
(Shanghai, People's Republic of China) 
and the Thousand-Youth-Talent Program of People's Republic of China.


\vspace{0.5cm}
\noindent{\emph{\bf Additional information }} 
The authors declare no competing financial interests.

\end{document}